\documentclass[prb,twocolumn,amsmath,amssymb,superscriptaddress,floatfix,nofootinbib]{revtex4-2}

\usepackage{graphicx}

\usepackage{dcolumn}

\usepackage{bm}

\usepackage{ textcomp } 
\usepackage{color} 
\usepackage{amsmath} 
\usepackage{amssymb} 
\usepackage{xcolor} 
\usepackage[colorlinks = true, linkcolor = blue, urlcolor = red, citecolor = blue, anchorcolor = blue]{hyperref} 
\usepackage{easyReview} 
\usepackage{mathdots} 
\usepackage{float} 
\usepackage{lineno} 
\usepackage{todonotes} 
\usepackage{url} 
\usepackage{array} 
\usepackage{todonotes} 
\usepackage{xcolor,colortbl} 
 
\usepackage{lipsum, babel} 
\usepackage{bbold}

\definecolor{LightBlue}{rgb}{0.8,0.8,0.8} 
 
\newcommand{\lS}{{\lambda_{\rm SOC}}} 
\newcommand{\bk}{{\bf k}}

\usepackage[normalem]{ulem}

\begin{document}

\title{Spin singlet topological superconductivity in the attractive Rashba Hubbard model}

\date{\today}

\author{Peter Doak} \affiliation{Computational Sciences and Engineering Division, Oak Ridge National Laboratory, Oak Ridge, Tennessee 37831, USA}

\author{Giovanni Balduzzi} \affiliation{Institute for Theoretical Physics, ETH Zurich, 8093 Zurich, Switzerland}

\author{Pontus Laurell} \affiliation{Department of Physics and Astronomy, University of Tennessee, Knoxville, Tennessee 37996, USA} \address{Computational Sciences and Engineering Division, Oak Ridge National Laboratory, Oak Ridge, Tennessee 37831, USA}

\author{Elbio Dagotto} \affiliation{Department of Physics and Astronomy, University of Tennessee, Knoxville, Tennessee 37996, USA} \affiliation{Materials Science and Technology Division, Oak Ridge National Laboratory, Oak Ridge, Tennessee 37831, USA}

\author{Thomas A. Maier} \affiliation{Computational Sciences and Engineering Division, Oak Ridge National Laboratory, Oak Ridge, Tennessee 37831, USA}

\begin{abstract}
	Fully gapped, spin singlet superconductors with antisymmetric spin-orbit coupling in a Zeeman magnetic field provide a promising route to realize superconducting states with non-Abelian topological order and therefore fault-tolerant quantum computation. Here we use a quantum Monte Carlo dynamical cluster approximation to study the superconducting properties of a doped two-dimensional attractive Hubbard model with Rashba spin-orbit coupling in a Zeeman magnetic field. We generally find that the Rashba coupling has a beneficial effect towards $s$-wave superconductivity.
	In the presence of a finite Zeeman field, when superconductivity is suppressed by Pauli pair-breaking, the Rashba coupling counteracts the spin imbalance created by the Zeeman field by mixing the spins, and thus restores superconductivity at finite temperatures. We show that this favorable effect of the spin-orbit coupling is traced to a spin-flip driven enhancement of the amplitude for the propagation of a pair of electrons in time-reversed states. Moreover, by inspecting the Fermi surface of the interacting model, we show that for sufficiently large Rashba coupling and Zeeman field, the superconducting state is expected to be topologically non-trivial.   
\end{abstract}

\maketitle

\section*{Introduction} 
In topological superconductors, the non-trivial topology of the bulk electronic structure leads to the emergence of Majorana bound states within the bulk superconducting gap \cite{Sato2017,Qi2011,Alicea2012}. These quasiparticles may be used for fault-tolerant quantum computing \cite{KITAEV20032}, and the search for new topological superconductors that host robust Majorana modes has therefore been an important priority but also a central challenge in quantum materials research. While topological superconductivity is usually associated with odd-parity spin triplet pairing, it was shown that spin-singlet, even parity superconductors can also host a non-Abelian topological phase in the presence of spin-orbit coupling and a Zeeman magnetic field \cite{Sato2009,Sato2010}. Experimental platforms to realize such a system include heterstructures of a semiconducting thin film sandwiched between an $s$-wave superconductor and a ferromagnetic insulator \cite{sau_generic_2010}, a two-dimensional electron gas adjacent to an interdigitated superconductor/ferromagnet structure \cite{lee_electrical_2012}, electric double layer transistors with an $s$-wave superconductor/ferromagnet heterostructure \cite{nagai_critical_2016}, and superfluids of cold atoms \cite{zhang_p_xip_y_2008,Sato2009}

Realizing topological superconductivity requires an intricate cooperation between helical states created by spin-orbital coupling, time-reversal symmetry breaking and superconductivity. Most studies of these ingredients, however, have used Bogoliubov-de-Gennes (BdG) weak-coupling mean-field theory \cite{Sato2009,Sato2010b,Daido2016,Yoshida2016}, which assumes that superconductivity is present and unaffected by the correlations, the spin-orbit coupling, or the Zeeman field. However, in order to provide general guiding principles for the design of topological superconducting materials, the effects of correlations, spin-orbit coupling and magnetic fields have to be treated on the same footing on a microscopic, beyond weak-coupling mean-field level, in order to properly assess the interplay between strong correlations and topology. Such work, however, is scarce, with only a few exceptions that include the dynamical mean-field theory (DMFT) work by Nagai \emph{et al.} \cite{nagai_critical_2016} and Lu \emph{et al.} \cite{lu_parity-mixing_2018}. Here we investigate these effects and the feedback between them on a microscopic level within numerical quantum Monte Carlo (QMC) dynamical cluster approximation (DCA) calculations of a Rashba-Hubbard model using a large enough cluster that properly accounts for the effects of the non-local Rashba coupling. We also use additional density matrix renormalization group (DMRG) calculations on a two-leg ladder (reported in the supplemental material) to show that our DCA results are robust. 

\section*{Model and Methods} 

We consider a two-dimensional square lattice attractive Rashba-Hubbard model in a Zeeman magnetic field. Its Hamiltonian is given by 
\begin{eqnarray}
	H &=& \sum_\bk \psi_\bk^\dagger (\epsilon_\bk\mathbb{1}-h\sigma_3+2\lS\,{\bm \sigma}\cdot {\bf g_k})\psi_\bk^{\phantom\dagger} \nonumber\\
	&+& U\sum_i n_{i\uparrow}n_{i\downarrow}\,. 
\label{eq:H} \end{eqnarray}
Here, we have used a spinor notation $\psi^\dagger_\bk=(c_{\bk\uparrow}^\dagger, c_{\bk\downarrow}^\dagger)$, with $c^\dagger_{\bk\sigma}$ creating an electron with wavevector $\bk$ and spin $\sigma = \uparrow, \downarrow$. For the square lattice with only nearest-neighbor hopping $t$, which we use as the energy unit ($t=1$), the energy dispersion is $\epsilon_\bk = -2t(\cos k_x+\cos k_y)$. ${\bm \sigma} = (\sigma_1, \sigma_2, \sigma_3)$ are the Pauli matrices, $h$ the Zeeman magnetic field, and $\lS$ the Rashba spin-orbit coupling with $g_\bk = (-\sin k_y, \sin k_x, 0)$. The second part of the Hamiltonian describes the on-site attractive interaction $U<0$ with the density operators $n_{i\sigma}=c^\dagger_{i\sigma}c^{\phantom\dagger}_{i\sigma}$. Here we have set $U=-4t$ and the electron density $\langle n\rangle=0.25$.

\begin{figure}
	[ht!] 
	\includegraphics[width=0.3\textwidth]{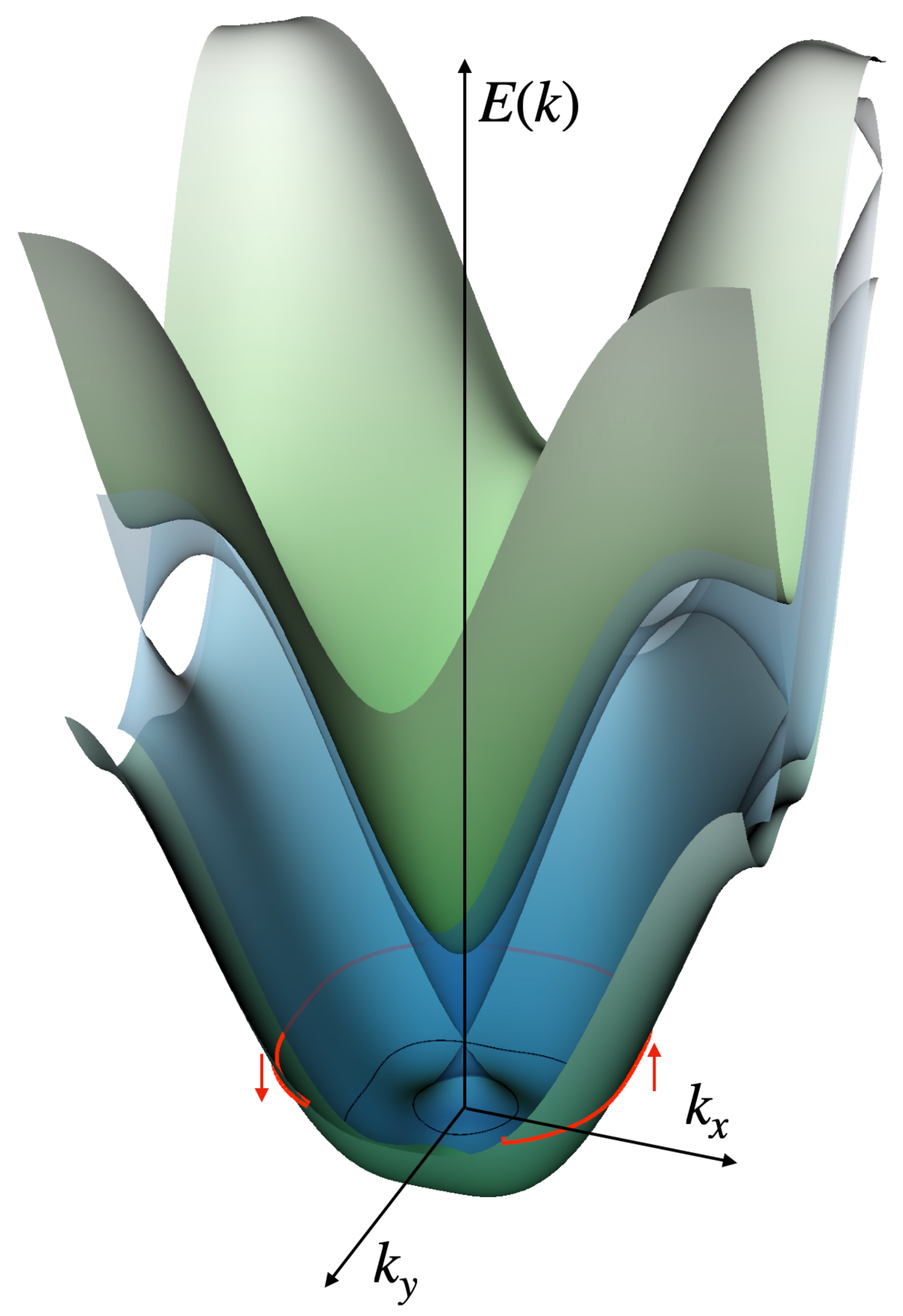} 
	\caption{\textbf{Illustration of the bandstructure with Rashba spin-orbit coupling and Zeeman field.} In the presence of finite Rashba spin-orbit coupling $\lS$, the Kramers degenerate bands for spin $\uparrow$ and $\downarrow$ split into two pseudospin bands, resulting in a Dirac cone near the $\Gamma$ point where the two bands touch (blue bandstructure). When the Zeeman field breaks time-reversal symmetry, a gap opens at $\Gamma$ (green bandstructure) resulting in a single ("spinless") Fermi surface (red line) when the chemical potential is tuned to an energy within the gap. \label{fig:1}} 
\end{figure}

The bandstructure $E(\bk)$ that results from diagonalizing the non-interacting part of the Hamilonian is schematically illustrated in Fig.~\ref{fig:1}. In the presence of a finite spin-orbit coupling $\lS$ but zero Zeeman field, the bandstructure splits into two pseudospin bands that are degenerate only at the $\Gamma$ point ($\bk=0$), resulting in a Dirac cone (blue surface). The Fermi surface always has two sheets, no matter where the chemical potential is located. For finite Zeeman field $h$, a gap opens at $\Gamma$ (green surface). In this case, when the chemical potential $\mu$ is tuned to fall within the gap, the Fermi surface consists of only a single (pseudospin) sheet with a helical spin structure, where the physical electron spin is pointing in opposite directions on opposite sides of the Fermi surface due to the spin-momentum locking induced by the spin-orbit coupling. This allows for the formation of spin-singlet Cooper pairs with opposite momenta $\bk, \uparrow$ and $-\bk, \downarrow$ in the presence of an attractive interaction. However, the pairing in this case is effectively spinless, since the other pseudospin degree of freedom is gapped out, and the superconducting state is expected to be topologically non-trivial. We note that while this pairing state is a spin-singlet pairing state in the original spin basis, it corresponds to a spin-triplet state in the pseudospin basis. In fact, Sato \emph{et al.} have shown that the $s$-wave BdG Hamiltonian can be mapped to a spin-less chiral $p$-wave superconductor in the chiral pseudospin basis \cite{Sato2010}. 

Here we use non-perturbative QMC/DCA \cite{hettler_nonlocal_1998,maier_quantum_2005,maier_pair_2011} calculations for the model in Eq.~(\ref{eq:H}) to study whether this physics can indeed be realized and to determine the superconducting transition temperature $T_c$ in the presence of both finite Rashba spin-orbit coupling $\lS$ and Zeeman field $h$. In the absence of these terms, the attractive Hubbard model in Eq.~(\ref{eq:H}) has been studied extensively over the past several decades \cite{blankenbecler_monte-carlo_1981,scalettar_phase_1989,micnas_superconductivity_1990,moreo_two-dimensional_1991,wilson_developments_2001,keller_dynamical_2001,paiva_critical_2004,paiva_fermions_2010,kaneko_bcsbec_2014,staar_two-particle_2014,fontenele_two-dimensional_2022}. Away from half-filling $\langle n\rangle=1$, this model has a finite-temperature Kosterlitz-Thouless $s$-wave superconducting transition that can be mapped out essentially exactly as QMC does not face a Fermion sign problem for this model \cite{paiva_critical_2004,staar_two-particle_2014,fontenele_two-dimensional_2022}.

When $\lS$ is finite but $h=0$, the model preserves time reversal symmetry. In this case, there is still no sign problem in the QMC \cite{wu_sufficient_2005}. For $|h|>0$, however, time reversal symmetry is broken, and the QMC calculations are not protected by symmetry from a sign problem (in this case a phase problem since the Hamiltonian is complex). The model with finite $\lS$ and $h$ was studied with single-site DMFT by Nagai \emph{et al.} in Ref.~\cite{nagai_critical_2016}. The single-site calculation is sign-problem free, but the purely non-local Rashba spin-orbit coupling vanishes in the effective single-site DMFT problem and is therefore not adequately treated in the calculation. Here, to properly take into account the effects of the Rashba coupling, we use an 8-site cluster for the DCA calculations and a continuous-time auxiliary-field QMC algorithm \cite{gull_continuous-time_2008,gull_submatrix_2011} to solve the effective cluster problem as implemented in the DCA++ code \cite{hahner_dca_2020}. The 8-site cluster is the smallest cluster for which the $\lS$ coupling does not vanish for the effective cluster problem. For the parameters of interest in this work, we do not find a strong phase problem for this cluster, 
allowing us to perform simulations for very low temperatures and accurately study the effects of $\lS$ and $h$ on the superconducting properties of the model. 

In order to do so, we calculate the $s$-wave pair-field susceptibility 
\begin{equation}
	P_s(T) = \int_0^\beta d\tau \langle {\cal T}_\tau \Delta^{\phantom\dagger}_s(\tau) \Delta_s^\dagger(0)\rangle\,, 
\label{eq:Ps} \end{equation}
with $\Delta^\dagger_s=1/\sqrt{N}\sum_\bk c^\dagger_{\bk\uparrow} c^\dagger_{-\bk\downarrow}$. The calculation of $P_s(T)$ within the DCA method is described in detail in the supplemental material. $T_c$ is determined as the temperature $T$ at which $P_s(T)$ diverges, or equivalently, $1/P_s(T)$ becomes zero (see supplemental material). We will also study its leading order term, the intrinsic pair-field susceptibility $P_{s,0}(T)$, which reflects the amplitude for the propagation of a pair of electrons in time-reversed momentum and spin states, in order to get insight into the pairing behavior. This quantity is given by 
\begin{equation}
	P_{s,0} = \frac{T}{N}\sum_k \left[G_{\uparrow\uparrow}(k)G_{\downarrow\downarrow}(-k) - G_{\uparrow\downarrow}(k)G_{\downarrow\uparrow}(-k)\right]\,.
\label{eq:Ps0} \end{equation}
Here we have used the notation $k=(\bk, i\omega_n)$ for fermionic Matsubara frequencies $\omega_n=(2n+1)\pi T$ and $G_{\sigma\sigma'}(k)$ is the fully interacting Green's function for the model in Eq.~\ref{eq:H}. The first term is the usual term for spin-singlet $(\bk\uparrow, -\bk\downarrow)$ pairs.
The second, spin-flip term $G_{\uparrow\downarrow}G_{\downarrow\uparrow}$ is only finite when the spin-orbit coupling $\lS$ mixes $\uparrow$ and $\downarrow$ spins, which leads to finite off-diagonal Green's function components $G_{\uparrow\downarrow}$ and $G_{\downarrow\uparrow}$.

\section*{Results and Discussion} 

\begin{figure}
	[ht] 
	\includegraphics[width=0.45\textwidth]{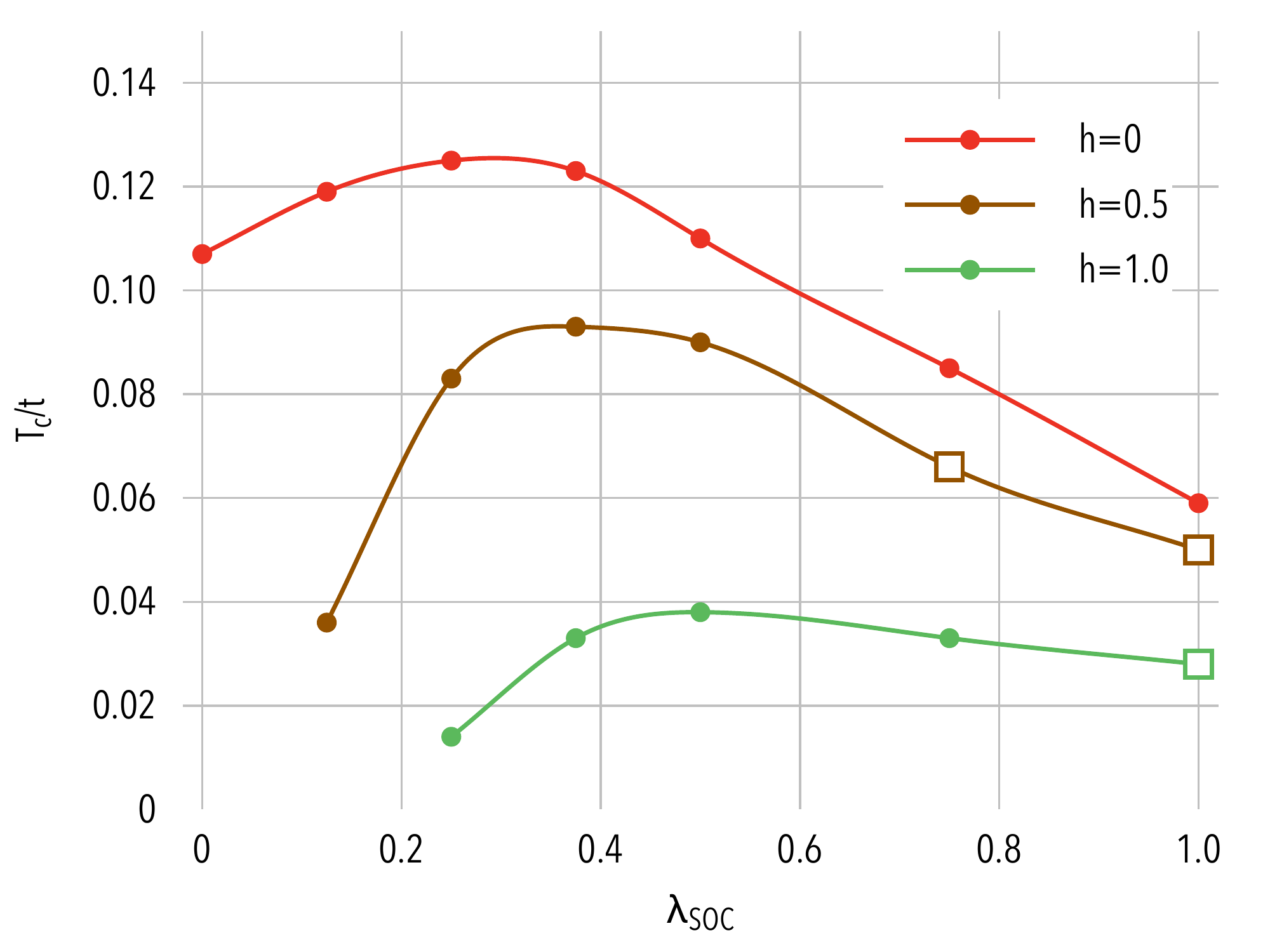} 
	\caption{\textbf{Superconducting transition in the attractive Rashba Hubbard model.} The superconducting transition temperature $T_c$ versus Rashba spin-orbit coupling $\lS$ for different Zeeman magnetic fields $h$. The data points with square symbols indicate the parameters for which the superconducting state is expected to be topologically non-trivial based on the results for the Fermi surface shown in Fig.~\ref{fig:4}. Results are shown for an electron filling $\langle n\rangle=0.25$ and $U=-4t$, and the DCA calculations were performed for an 8-site cluster. \label{fig:2} } 
\end{figure}

Fig.~\ref{fig:2} shows the superconducting transition temperature $T_c$  versus the Rashba spin-orbit coupling $\lS$ for different values of the Zeeman magnetic field $h$. $T_c$ was obtained as the temperature at which the $s$-wave pair-field susceptibility $P_s(T)$ diverges. We find non-monotonic behavior for all values of $h$. 
For $h=0$, $T_c\approx 0.11t$ and increases by about 20\% to $T_c\sim 0.13t$ at $\lS=0.25$ before it decreases at larger $\lS$. Similar behavior is found at finite $h$. In this case, superconductivity is suppressed for fields larger than the upper critical field when $\lS=0$ due to Pauli pair breaking. For $h=0.5$ $(1.0)$, we do not find a superconducting transition for $\lS=0$ (0 and 0.125). For larger $\lS$, however, superconductivity is restored and $T_c$ initially increases with $\lS$ before it decreases at larger $\lS$. This behavior is consistent with our DMRG calculations of a two-leg Rashba-Hubbard ladder as reported in the supplemental material, where we find that the binding energy for a pair of holes is negative and has a minimum at intermediate Rashba coupling for finite Zeeman fields. The DMRG calculations also show non-monotonic behavior in the $\lS$ dependence of the on-site pair-pair correlations with a pronounced enhancement for intermediate Rashba coupling. Our results are also consistent with those found previously in single-site DMFT calculations for this model \cite{nagai_critical_2016}. For the data points with the square symbols at large $\lS$ and finite $h$, we expect the superconducting state to be topologically non-trivial, based on results for the normal state Fermi surface, as we will discuss later. 

\begin{figure}
  [ht] 
  \includegraphics[width=0.45\textwidth]{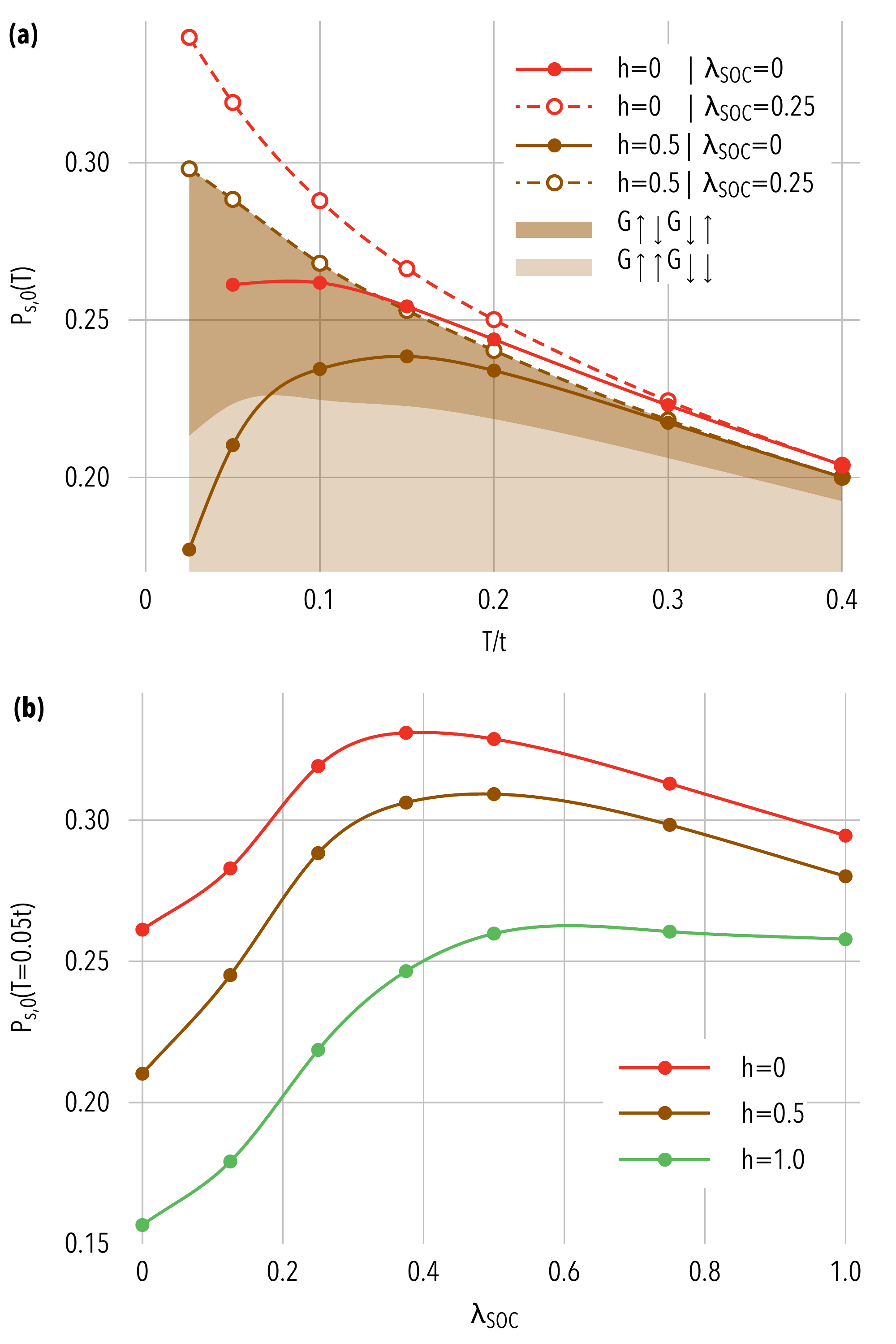} 
  \caption{\textbf{Intrinsic pair-field susceptibility.} Panel (a): Temperature dependence of the $s$-wave intrinsic pair-field susceptibility $P_{s,0}(T)$ for different Zeeman fields $h$ and Rashba couplings $\lS$. The light (dark) shaded regions show the contributions in Eq.~(\ref{eq:Ps0}) of the first $G_{\uparrow\uparrow}G_{\downarrow\downarrow}$ (second $G_{\uparrow\downarrow}G_{\downarrow\uparrow}$) terms to $P_{s,0}(T)$ for $h=0.5$, $\lS=0.25$.  Panel (b): $P_{s,0}(T)$ at fixed temperature $T=0.05t$ versus $\lS$. All results are shown for an electron filling $\langle n\rangle=0.25$ and $U=-4t$, and the DCA calculations were performed for an 8-site cluster. \label{fig:3} } 
\end{figure}

From Fig.~\ref{fig:2} it is clear that $\lS$ increases the Pauli-limit upper critical field above which superconductivity is suppressed, and thus restores superconductivity for fields above the upper critical field of the system with $\lS=0$. We now discuss results for the intrinsic $s$-wave pair-field susceptibility $P_{s,0}$ defined in Eq.~\ref{eq:Ps0} in order to provide insight into this behavior. In Fig.~\ref{fig:3}, we show results for the temperature and $\lS$ dependence of $P_{s,0}$. In conventional (BCS) theory, this quantity has a logarithmic (Cooper) divergence as $T\rightarrow 0$, so that any attractive interaction, no matter how weak, leads to a superconducting transition at finite temperature. In an unconventional superconductor, however, the physics can be different. In the cuprate pseudogap phase, for example, this Cooper instability is absent, and the superconducting transition is driven by an effective interaction that increases with decreasing temperature \cite{maier_pairing_2016}. Thus, a logarithmic divergence in $P_{s,0}(T)$ is a sufficient, but not necessary condition for a superconducting instability to occur.   

Panel $\bf{(a)}$ in Fig.~\ref{fig:3} plots the temperature $T$ dependence of $P_{s,0}(T)$ for two different values of $h$ and $\lS$. For $\lS=0$ (solid circles), finite $h=0.5$ splits the Kramers degenerate Fermi surface into two ($\uparrow$ and $\downarrow$) sheets. As a result, there are no states available at $-\bk$ on the $\downarrow$ sheet to pair with the $\bk, \uparrow$ state. Consequently, $P_{s,0}(T)$ is significantly suppressed at low temperatures by the Zeeman field. A finite $\lS$ mixes spin $\uparrow$ with spin $\downarrow$ and thus counteracts the spin imbalance created by the $h$ field. Consistent with this expectation, for $\lS=0.25$ (open cirlces), the low temperature behavior of $P_{s,0}(T)$ changes significantly, now showing a strong upturn with decreasing temperature even for $h>0$, which eventually leads to the finite $T_c$ shown in Fig.~\ref{fig:2}. Albeit less dramatic, this enhancement occurs even for $h=0$ and is the reason for the enhancement of $T_c$ with finite spin-orbit coupling $\lS$. For $h=0.5$ and $\lS=0.25$, the two terms, $G_{\uparrow\uparrow}G_{\downarrow\downarrow}$ and $G_{\uparrow\downarrow}G_{\downarrow\uparrow}$ that contribute to $P_{s,0}$ in Eq.~(\ref{eq:Ps0}) are shown as shaded regions. The standard $G_{\uparrow\uparrow}G_{\downarrow\downarrow}$ contribution remains suppressed at low temperatures, even in the presence of finite $\lS$. In contrast, the spin-flip $G_{\uparrow\downarrow}G_{\downarrow\uparrow}$ contribution keeps rising with decreasing temperature, leading to the low-temperature increase of $P_{s,0}(T)$ for finite $\lS$. The $\lS$ and $h$ dependence of $P_{s,0}(T)$ at fixed $T=0.05t$ in panel \textbf{(b)} closely tracks the $\lS$ dependence of $T_c$ in Fig.~\ref{fig:2}, showing that it is indeed the effect of $\lS$ and $h$ on the intrinsic pair-field susceptibility $P_{s,0}$ that determines $T_c$. 

\begin{figure*}
  [t!] 
  \includegraphics[width=0.95\textwidth]{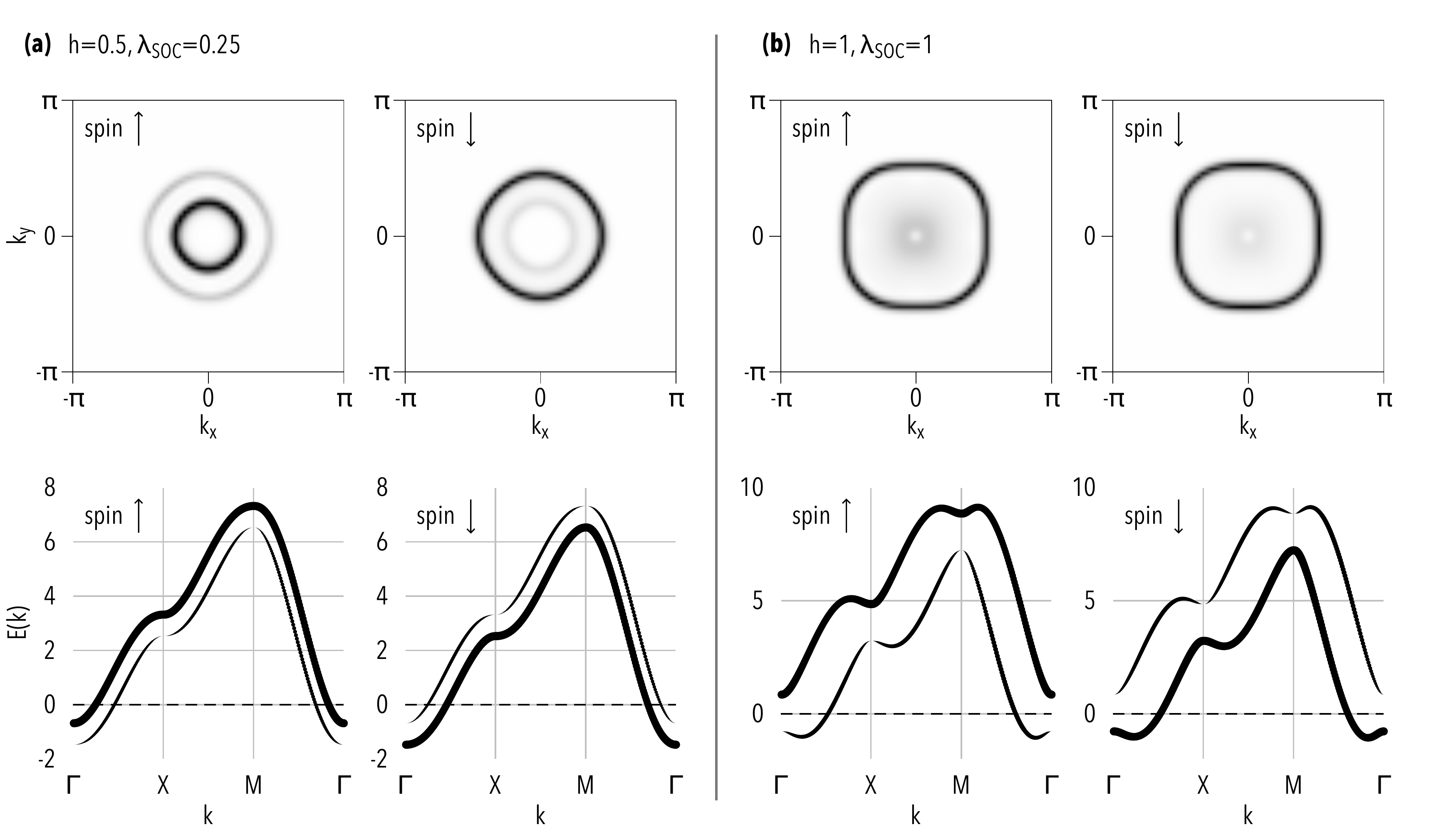} 
  \caption{\textbf{Fermi surface and bandstructure.} Top panels: The gradient of the momentum space occupation, $|\nabla_k n(k)|$ for (a) $h=0.5$, $\lS=0.5$ and (b) $h=1.0$, $\lS=1.0$ for spin $\uparrow$ and $\downarrow$. Bottom panels: Corresponding bandstructure in Hartree approximation with the weight of spin $\uparrow$ and $\downarrow$ indicated by line thickness. \label{fig:4}} 
\end{figure*}

The key for understanding the different effects of $\lS$ and $h$ on $P_{s,0}(T)$ is the Fermi surface and bandstructure plotted in Fig.~\ref{fig:4}. Here we show two different parameter sets: $h=0.5$, $\lS=0.25$ in panel \textbf{(a)} and $h=1$, $\lS=1$ in panel \textbf{(b)}. The top two panels show $|\nabla_\bk n^\sigma_\bk|$ as a proxy for the Fermi surface, where $n^\sigma_\bk=c^\dagger_{\bk\sigma}c^{\phantom\dagger}_{\bk\sigma}$ is the occupation in momentum space for spin $\sigma$. The bottom panels show the bandstructure of the non-interacting model, but including the Hartree term of the self-energy, to serve as a leading order approximation of the interacting single-particle spectrum that can be compared with $|\nabla_\bk n^\sigma_\bk|$ in the top panels. 
As is clear from these plots, the Fermi level crossing of the bands and their spin $\sigma$ weights are consistent with the Fermi surface plots in the top panels.

For the case with $h=0.5$, $\lS=0.25$ in panel \textbf{(a)}, one sees two bands crossing the Fermi level and two Fermi surface sheets closed around the $\Gamma$ point with with very different spin weights. 
The spin-orbit induced admixture of the opposite spin, although weak, leads to a finite spin-flip $G_{\uparrow\downarrow}G_{\downarrow\uparrow}$ contribution to $P_{s,0}(T)$ that is immune to the field induced suppression at low temperature, thus restoring the superconducting instability. 

For the second case with $h=1$, $\lS=1$ in panel \textbf{b}, the situation is very different. The splitting of the bands is much larger resulting in a single Fermi surface sheet only. The states on this sheet have predominantly spin $\downarrow$ character, but now with a much larger admixture of spin $\uparrow$ electrons. For this case of a single sheet, the superconducting state is expected to be topologically non-trivial, since the pairing (in the chiral pseudospin basis) is effectively spinless due to the absence of the second Fermi surface sheet. As indicated by the square data points in Fig.~\ref{fig:2}, the cases with $h=0.5$ and $\lS \ge 0.75$ are also expected to be topologically non-trivial based on their Fermi surface (not shown).

\section*{Summary and Conclusions}

We have used a dynamic cluster quantum Monte Carlo approximation to study $s$-wave superconductivity in the attractive Hubbard model in the presence of a Rashba spin-orbit coupling $\lS$ and a Zeeman magnetic field $h$ for an electron filling $\langle n\rangle = 0.25$. Under certain conditions, these ingredients can lead to a spin-singlet superconducting state that is topologically non-trivial. We have found that a Rashba coupling with moderate strength $\lS\sim t$ has a beneficial effect towards superconductivity. By mixing spin $\uparrow$ with $\downarrow$ states, it counteracts the spin imbalance generated by the $h$ field and thus creates $-{\bf k}_F,\downarrow$ Fermi level states that can pair with degenerate ${\bf k}_F,\uparrow$ states. This promotes a superconducting transition for fields well above the Pauli limit upper critical field at $\lS=0$. We show that this favorable effect of $\lS$ is traced to a spin-flip driven enhancement of the electron pair propagation amplitude $P_{s,0}$, which is induced by the spin mixture. Finally, we used the gradient of the momentum space occupation $n(\bk)$ to obtain information on the Fermi surface of the interacting system for different $\lS$ and $h$. For sufficiently large $\lS$ and $h$, we find that the Fermi level falls within the gap of the effective two-band system, and the Fermi surface consists of only a single (pseudospin) sheet. For this case, the superconducting state below $T_c$ is effectively spin-less and therefore expected to be topologically non-trivial. These results give new insight into the effects of spin-orbit coupling and magnetic fields on the superconducting behavior of correlated electron systems, and thus provide general guidance on how to tune the relative strengths of these couplings in the search for new topological superconductors.

\begin{acknowledgements}
We acknowledge useful discussions with Fakher Assaad. This work was supported by the U.S. Department of Energy, Office of Science, Basic Energy Sciences, Materials Sciences and Engineering Division. P.D. and G.B. acknowledge support from the Scientific Discovery through Advanced Computing (SciDAC) program funded by the U.S. Department of Energy, Office of Science, Advanced Scientific Computing Research and Basic Energy Sciences, Division of Materials Sciences and Engineering for code development. An award of computer time was provided by the INCITE program. This research also used resources of the Oak Ridge Leadership Computing Facility, which is a DOE Office of Science User Facility supported under Contract DE-AC05-00OR22725. This manuscript has been authored by UT-Battelle, LLC, under Contract No. DE-AC0500OR22725 with the U.S. Department of Energy.  The United States Government retains and the publisher, by accepting the article for publication, acknowledges that the United States Government retains a nonexclusive, paid-up, irrevocable, world-wide license to publish or reproduce the published form of this manuscript, or allow others to do so, for the United States Government purposes. The Department of Energy will provide public access to these results of federally sponsored research in accordance with the DOE Public Access Plan (\url{http://energy.gov/downloads/doe-public-access-plan}). 
\end{acknowledgements}

\bibliography{main}

\end{document}


\title{Supplemental Material: Spin singlet topological superconductivity in the attractive Rashba Hubbard model}

\date{\today}

\author{Peter Doak} \affiliation{Computational Sciences and Engineering Division, Oak Ridge National Laboratory, Oak Ridge, Tennessee 37831, USA}

\author{Giovanni Balduzzi} \affiliation{Institute for Theoretical Physics, ETH Zurich, 8093 Zurich, Switzerland}

\author{Pontus Laurell} \affiliation{Department of Physics and Astronomy, University of Tennessee, Knoxville, Tennessee 37996, USA} \address{Computational Sciences and Engineering Division, Oak Ridge National Laboratory, Oak Ridge, Tennessee 37831, USA}

\author{Elbio Dagotto} \affiliation{Department of Physics and Astronomy, University of Tennessee, Knoxville, Tennessee 37996, USA} \affiliation{Materials Science and Technology Division, Oak Ridge National Laboratory, Oak Ridge, Tennessee 37831, USA}

\author{Thomas A. Maier} \affiliation{Computational Sciences and Engineering Division, Oak Ridge National Laboratory, Oak Ridge, Tennessee 37831, USA }

\date{\today}

\flushbottom
\maketitle





\section{DCA calculation of the s-wave pair-field susceptibility}

In order to calculate the $s$-wave pair-field susceptibility (Eq.~(2) in the main text) for the model in Eq.~(1), we follow the usual DCA formalism described in Refs.~\cite{jarrell_quantum_2001, maier_quantum_2005} to calculate susceptibilities for the lattice in the thermodynamic limit. This requires a calculation of the 4-point two-particle Green's function
\begin{eqnarray}
    \label{eq:G2}
    G_{2, \sigma_1\dots\sigma_4}(x_1,x_2;x_3,x_4)=\langle T_\tau c^{\phantom\dagger}_{\sigma_1}(x_1)c^{\phantom\dagger}_{\sigma_2}(x_2)c^{\dagger}_{\sigma_3}(x_3)c^{\dagger}_{\sigma_4}(x_4)\rangle\,,
\end{eqnarray}
where the combined index $x_i=({\bf x}_i,\tau_i)$ has both spatial, ${\bf x}_i$, and imaginary time, $\tau_i$ coordinates. Fourier-transforming on both the space and time variables gives $G_{2,\sigma_1\dots\sigma_4}(k_1, k_2; k_3, k_4)$ with $k_\ell=({\bf k}_\ell, i\omega_{n_\ell})$. The $s$-wave pair-field susceptibility $P_s(T)$ defined in Eq.~(2) in the main text is then obtained from
\begin{eqnarray}
    \label{eq:Ps}
    P_s(T) = \frac{T}{N}\sum_{k,k'}G_{2,\uparrow\downarrow\uparrow\downarrow}(k, -k, k', -k')\,.
\end{eqnarray}

\begin{figure}
    [ht!] 
    \includegraphics[width=1\textwidth]{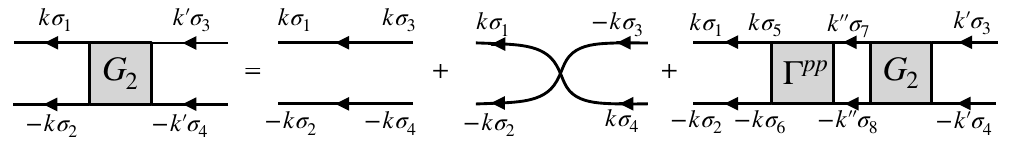} 
    \caption{\textbf{Diagrams for the Bethe-Salpeter equation in the particle-particle channel.} For finite (spin-mixing) Rashba coupling, an additional diagram with crossing Green's function legs contributes in leading order. This diagram gives rise to the enhancement of $s$-wave pair correlations found at low temperatures. \label{fig:S1}} 
\end{figure}

The two-particle Green's function $G_s$ for the lattice in the thermodynamic limit is obtained from the Bethe-Salpeter equation in the particle-particle channel shown diagrammatically in Fig.~\ref{fig:S1},
\begin{eqnarray}
    \label{eq:BSE}
    G_{2,\sigma_1\dots\sigma_4}(k, -k, k', -k') &=& G_{\sigma_1\sigma_3}(k)G_{\sigma_2\sigma_4}(-k)\delta_{k,k'} - G_{\sigma_1\sigma_4}(k)G_{\sigma_2\sigma_3}(-k)\delta_{k,-k'} \\\nonumber
    &+& \frac{T}{N}\sum_{k''}\sum_{\sigma_5\dots\sigma_8}G_{\sigma_1\sigma_5}(k)G_{\sigma_2\sigma_6}(-k)\Gamma^{pp}_{\sigma_5\dots\sigma_8}(k,-k,k'',-k'')G_{2,\sigma_7\sigma_8\sigma_3\sigma_4}(k'',-k'',k',-k')
\end{eqnarray}
Here, $G_{\sigma\sigma'}(k)$ is the single-particle Green's function, which, due to the Rashba spin-orbit coupling is off-diagonal in the spin, and $\Gamma^{pp}_{\sigma_4\dots\sigma_1}(k, -k, k', -k')$ is the irreducible particle-particle vertex. In the DCA, its momentum dependence is reduced to that of the effective cluster problem, i.e. $\Gamma^{pp}_{\sigma_4\dots\sigma_1}(k, -k, k', -k') \approx \Gamma^{pp}_{\sigma_4\dots\sigma_1}(K, -K, K', -K')$, where $K = ({\bf K}, i\omega_n)$ contains the cluster momenta ${\bf K}$. Using the cluster vertex $\Gamma^{pp}_{\sigma_4\dots\sigma_1}(K, -K, K', -K')$, one can then calculate the coarse-grained two-particle Green's function for the lattice
\begin{eqnarray}
    \label{eq:BSEcg}
    {\bar G}_{2,\sigma_1\dots\sigma_4}(K, -K, K', -K') &\equiv& \frac{N_c^2}{N^2}\sum_{{\bf k}\in {\cal P}_{\bf K}}\sum_{{\bf k'}\in {\cal P}_{\bf K'}}G_{2,\sigma_1\dots\sigma_4}(k, -k, k', -k')\\
    &=& \bar{G}_{2,\sigma_1\dots\sigma_4}^0(K, -K, K', -K')+\frac{T}{N_c}\sum_{K''}\sum_{\sigma_5\dots\sigma_8}\bar{G}_{2,\sigma_1\dots\sigma_4}^{d}(K, -K, -K, K)\times \nonumber\\
    &\times&\Gamma^{pp}_{\sigma_5\dots\sigma_8}(K,-K,K'',-K''){\bar G}_{2,\sigma_7\sigma_8\sigma_3\sigma_4}(K'', -K'', K', -K')\,.
\end{eqnarray}
Here, the sums over ${\bf k}$, ${\bf k'}$ and ${\bf k''}$ have been partially carried out over the $N_c$ DCA coarse-graining patches ${\cal P}_{\bf K}$, etc. \cite{maier_quantum_2005}, with $N_c$ the cluster size, so that all quantities now depend on the cluster momenta ${\bf K}$ only. The coarse-grained bare propagators
\begin{eqnarray}
    \bar{G}_{2,\sigma_1\dots\sigma_4}^0(K,-K,K',-K') = \frac{N_c}{N}\sum_{{\bf k}\in{\cal P}_{\bf K}}[G_{\sigma_1\sigma_3}(k)G_{\sigma_2\sigma_4}(-k)]\,\delta_{K,K'} - \frac{N_c}{N}\sum_{{\bf k}\in{\cal P}_{\bf K}} [G_{\sigma_1\sigma_4}(k)G_{\sigma_2\sigma_3}(-k)]\,\delta_{K,-K'}
\end{eqnarray}
and 
\begin{eqnarray}
    \bar{G}_{2,\sigma_1\dots\sigma_4}^{d}(K, -K, K', -K') = \frac{N_c}{N}\sum_{{\bf k}\in{\cal P}_{\bf K}} [G_{\sigma_1\sigma_3}(k)G_{\sigma_2\sigma_4}(-k)]\, \delta_{K, -K'}
\end{eqnarray}
only has the (diagonal) first term. By including the spin variables in the combined indices $K=({\bf K}, \omega_n, \sigma_1, \sigma_2)$ and $K'=({\bf K'}, \omega_{n'}, \sigma_3, \sigma_4)$, this equation can be conveniently written in matrix form (in $K$ and $K'$)
\begin{eqnarray}
    \label{eq:BSEM}
    {\bf \bar G}_2 = {\bf \bar G}_2^0 + {\bf \bar G}_2^{d}{\bf \Gamma}^{pp}{\bf \bar G}_2\,.
\end{eqnarray}
The cluster vertex ${\bf \Gamma}^{pp}$ is determined from an analogous equation for the cluster two-particle Green's function $G_2^c$ \cite{maier_quantum_2005}
\begin{eqnarray}
    \label{eq:BSECluster}
    {\bf G}^c_2 = {\bf G}^{c,0}_2 + {\bf G}^{c,d}_2{\bf \Gamma}^{pp}{\bf G}^c_2\,.
\end{eqnarray}
The bare cluster propagators $G_{2,\sigma_1\dots\sigma_4}^{c,0}(K, -K, K', -K') = G^c_{\sigma_1\sigma_3}(K)G^c_{\sigma_2\sigma_4}(-K)\delta_{K,K'} - G^c_{\sigma_1\sigma_4}(K)G^c_{\sigma_2\sigma_3}(-K)\delta_{K,-K'}$, and $G_2^{c,d}(K) = G^c_{\sigma_1\sigma_3}(K)G^c_{\sigma_2\sigma_4}(-K)\delta_{K,K'}$, where $G^c_{\sigma\sigma'}(K)$ is the single-particle cluster Green's function. 

\begin{figure}
    [ht!] 
    \includegraphics[width=0.7\textwidth]{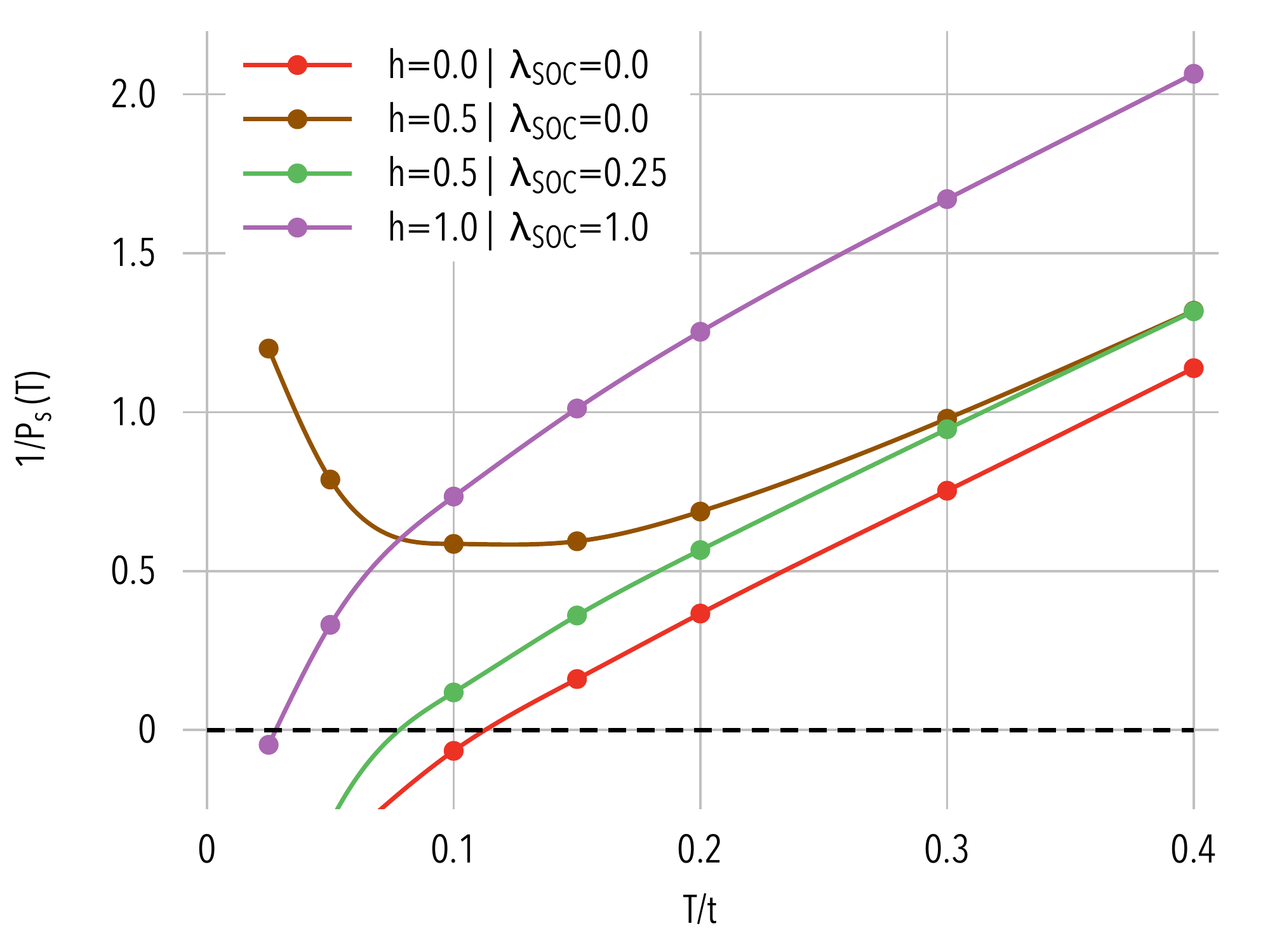} 
    \caption{\textbf{Temperature dependence of inverse $s$-wave pair-field susceptibility.} The inverse $s$-wave pair-field susceptibility $1/P_s(T)$ calculated according to Eq.~(\ref{eq:Ps}) versus temperature for different combinations of magnetic field $h$ and Rashba spin-orbit coupling $\lambda_{SOC}$. The values for the superconducting transition temperature $T_c$ shown in Fig.~1 in the main text are obtained from the temperature where $1/P_s(T)$ crosses zero (dashed line). \label{fig:S2}} 
\end{figure}

The extraction of the cluster vertex ${\bf \Gamma}^{pp}$ from the Bethe-Salpeter equation (\ref{eq:BSECluster}) involves an inversion of ${\bf G}_2^c$ and ${\bf G}_2^{c,0}$. The addition of the second term with crossed Green's function legs proportional to $\delta_{K,-K'}$ makes these matrices singular.  However, after some matrix arithmetics, an equation can be obtained for ${\bf \bar G}_2$ that does not involve the inversion of these matrices \cite{nagai_critical_2016}
\begin{eqnarray}
    \label{eq:G2Final}
    {\bf \bar G}_2 = {\bf G}_2^c \left[ \left( [{\bf \bar G}_2^{(1)}]^{-1} - [{\bf G}_2^{c,(1)}]^{-1}\right){\bf G}_2^c + {\bf B}^c\right]^{-1}{\bf B}\,.
\end{eqnarray}
Here ${\bf B} = [{\bf \bar G}_2^{d}]^{-1}{\bf \bar G}_2^{0}$ and the corresponding cluster quantity ${\bf B}^c = [{\bf G}_2^{c,d}]^{-1}{\bf G}_2^{c,0}$.
The lattice $s$-wave pair-field susceptibility $P_s(T)$ is then obtained from ${\bf \bar G}_2$ as
\begin{eqnarray}
    \label{eq:Ps}
    P_s(T) = \frac{T}{N_c}\sum_{K,K'}{\bar G}_{2,\uparrow\downarrow\uparrow\downarrow}(K,-K,K',-K')\,,
\end{eqnarray}
and the intrinsic pair-field susceptibility $P_{s,0}(T)$ in Eq.~(3) in the main text from
\begin{eqnarray}
    P_{s,0}(T)=\frac{T}{N_c}\sum_{K,K'} {\bar G}_{2,\uparrow\downarrow\uparrow\downarrow}^0(K,-K,K',-K')\,.
\end{eqnarray}
The temperature dependence of the inverse of $P_s(T)$ is shown for a selected set of parameters in Fig.~\ref{fig:S2}.

\section{Density matrix renormalization group analysis of an attractive Rashba-Hubbard two-leg ladder}

In order to investigate the robustness of the DCA results, we also carry out complementary density matrix renormalization group (DMRG) \cite{PhysRevLett.69.2863, PhysRevB.48.10345} calculations on a two-leg ladder with open boundaries using the DMRG++ software \cite{Alvarez2009}, working at zero temperature. This provides accurate insight into the real-space behavior of a minimal version of the 2D problem \cite{PhysRevB.45.5744, doi:10.1126/science.271.5249.618}. We first rewrite the hopping part of Eq.~(1) in real space as,
\begin{align}
    H_0 &=  -t\sum_{\langle i,j\rangle,\sigma} \left[ c_{i\sigma}^\dagger c_{j\sigma} + \mathrm{H.c.} \right] -h \sum_i \left( n_{i\uparrow} - n_{i\downarrow} \right) \nonumber\\
        &+  2\lambda_\mathrm{SOC} \sum_{\langle i,j\rangle} \sum_{\sigma,\sigma'} c_{i\sigma}^\dagger \left[ \alpha_{ij}^x \sigma_{\sigma\sigma'}^y - \alpha_{ij}^y \sigma_{\sigma\sigma'}^x  \right]c_{j\sigma'},
\end{align}
where $\langle \dots\rangle$ denotes summation over nearest neighbors, $\alpha_{ij}^\mu \equiv i \left( \delta_{i,j+a_\mu} - \delta_{i,j-a_\mu}\right)$, $a_\mu$ denotes translation in the $\mu$ direction \cite{PhysRevB.101.235149}, and we take $\hat{x}$ ($\hat{y}$) to be the long (short) direction of the ladder, namely along legs (across rungs). Here we consider ladders up to length $L_x=48$ (and width $L_y=2$). In obtaining the ground states, a truncation error below $10^{-9}$ was targeted and obtained for all parameters by keeping up to $m=1500$ states. Explicit reorthogonalization was used at each step.

\begin{figure}[bt!]
    \includegraphics[width=0.7\columnwidth]{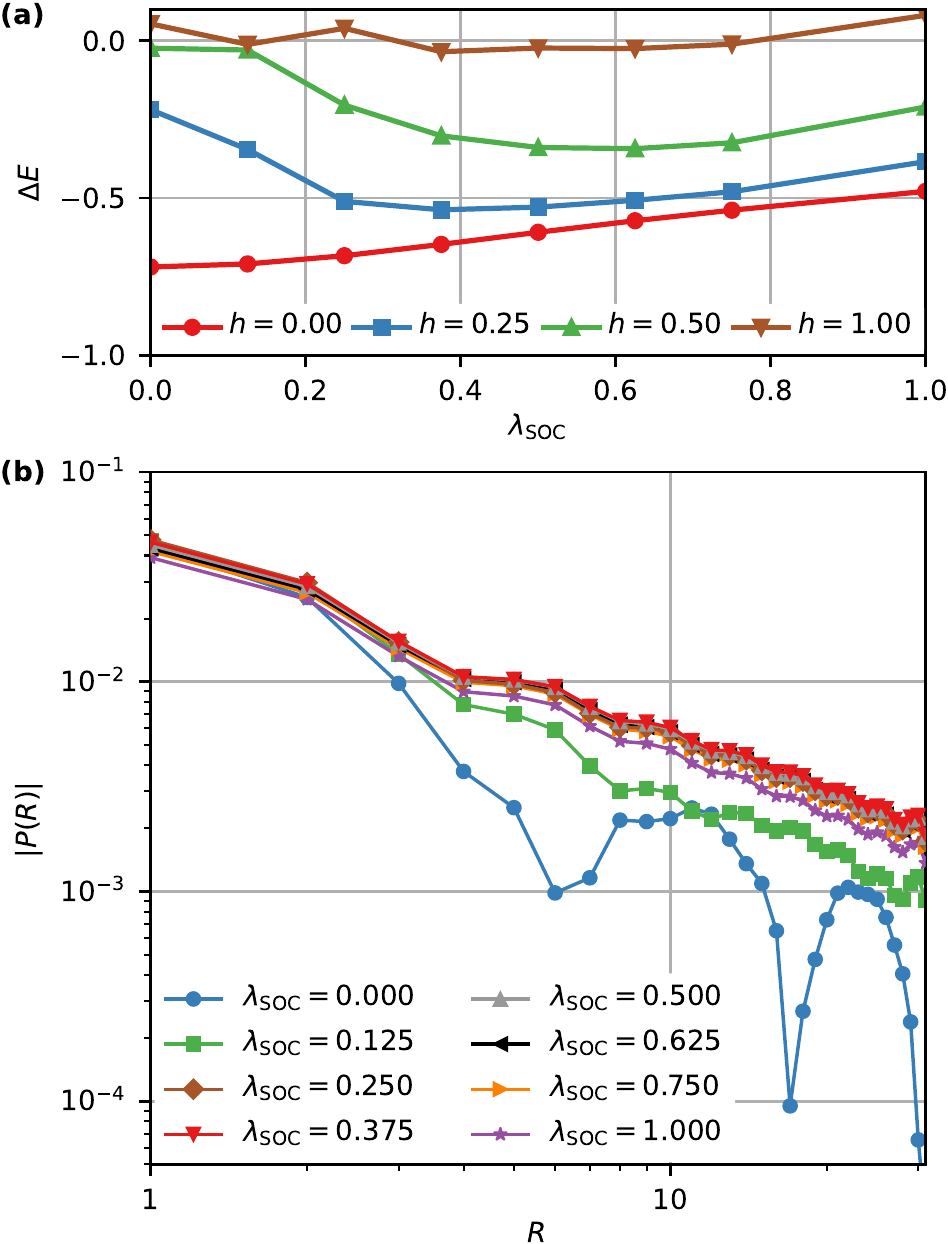}
    \caption{\textbf{Rashba-Hubbard two-leg ladder.} Panel (a): binding energy for $L_x=32$. Panel (b): on-site singlet pair-pair correlations for $L_x=48$ and $h=0.5$. Results in both panels are obtained using DMRG for electron filling $\langle n\rangle =0.25$ and $U=-4t$.}
    \label{fig:ladder}
\end{figure}
The first quantity of interest is the binding energy \cite{RevModPhys.66.763},
\begin{equation}
\Delta E \left(N\right) =   E_0 \left(N-2\right) + E_0 \left(N\right) -2E_0 \left(N-1\right),
\end{equation}
where $E_0 \left(N\right)$ denotes the ground state energy of the system with $N$ electrons present. $\Delta E \left(N\right)<0$ indicates it is favorable for two holes to form a Cooper pair bound state, a requirement for pairing to occur. $\Delta E= 0$ for two independent holes, but $\Delta E>0$ can also occur due to finite-size effects. The binding energy for $L_x=32$ is plotted for several fields and $\lambda_\mathrm{SOC}$ values in Fig.~\ref{fig:ladder}(a). For intermediate fields $0.25\leq h\leq 0.5$ we find a minimum in the binding energy at intermediate Rashba coupling, which reflects an enhanced tendency to pairing and is in qualitative agreement with the DCA results. At $h=0.5$ the value of the binding energy only becomes appreciably negative at finite $\lambda_\mathrm{SOC}\geq 0.25$. At zero field, however, the trend is monotonous in $\lambda_\mathrm{SOC}$. 
Some such differences between the DMRG and DCA results may be expected due to the difference in dimensionality, but the qualitative agreement at $h=0.5$ indicates the trend is general and robust.

We next consider pair-pair correlations for $h=0.5$. We focus on on-site singlet pairs, which are favored by the attractive Hubbard interaction. The pair creation operator on rung $i$ and leg $a$ can be written $    S_\mathrm{on-site}^{\dagger}\left( i\right) =    c_{ia\uparrow}^\dagger c_{ia\downarrow}^\dagger$ and the corresponding correlation function is given by
\begin{equation}
    P(R)    = \frac{1}{N_R} \sum_i \left\langle S_\mathrm{on-site}^{\dagger} (i) S_\mathrm{on-site}\phantom{} (i+R) \right\rangle,  \label{eq:pairpair:singlet}
\end{equation}
where $N_R$ denotes the number of total neighbors at distance $R$ from site $i$, summed over all sites. We neglect eight rungs at each end of the ladder in order to minimize edge effects. Correlations for a ladder with $L_x=48$ are plotted in Fig.~\ref{fig:ladder}(b), showing power-law behavior as expected for a quasi-1D system. We see a pronounced enhancement of the correlation function upon introduction of the Rashba coupling, with a maximum near $\lambda_\mathrm{SOC}=0.375$. The non-monotonous behavior matches that of the binding energy in Fig.~\ref{fig:ladder}(a), and is consistent with the behavior of the $s$-wave intrinsic pair-field susceptibility of the 2D system shown in Fig.~3. The beneficial effect of Rashba spin-mixing on singlet pairing becomes evident by comparing the correlations at finite $\lambda_\mathrm{SOC}$ with the much weaker correlations at $\lambda_\mathrm{SOC}=0$.

\bibliography{suplbib}

\bibliographystyle{apsrev4-2}

\noindent